# Discovery of 2D Materials via Symmetry-Constrained Diffusion Model

Shihang Xu[1,2], Shibing Chu[1,2*], Rami Mrad[1,2], Zhejun Zhang[1,2], Zhelin Li[1,2], Runxian Jiao[1,2], Yuanping Chen[1,2*]

1  School of Physics and Electronic Engineering, Jiangsu University, Zhenjiang, Jiangsu 212013, PR China

2 Jiangsu Engineering Research Center on Quantum Perception and Intelligent Detection of Agricultural Information, Zhenjiang, 212013, China

Corresponding author: c@ujs.edu.cn (S.C.), chenyp@ujs.edu.cn (Y. C.)

## ABSTRACT

Generative model for 2D materials has shown significant promise in accelerating the material discovery process. The stability and performance of these materials are strongly influenced by their underlying symmetry. However, existing generative models for 2D materials often neglect symmetry constraints, which limits both the diversity and quality of the generated structures. Here, we introduce a symmetry-constrained diffusion model (SCDM) that integrates space group symmetry into the generative process. By incorporating Wyckoff positions, the model ensures adherence to symmetry principles, leading to the generation of 2,000 candidate structures. DFT calculations were conducted to evaluate the convex hull energies of these structures after structural relaxation. From the generated samples, 843 materials that met the energy stability criteria ($E_{\text{hull}}$ < 0.6 eV/atom) were identified. Among these, six candidates were selected for further stability analysis, including phonon band structure evaluations and electronic properties investigations, all of which exhibited phonon spectrum stability. To benchmark the performance of SCDM, a symmetry-unconstrained diffusion model was also evaluated via crystal structure prediction model. The results highlight that incorporating symmetry constraints enhances the effectiveness of generated 2D materials, making a contribution to the discovery of 2D materials through generative modeling.

# 1. Introduction

Since the discovery and synthesis of graphene, two-dimensional (2D) materials have attracted considerable attention in scientific research.[1] Owing to their distinctive layered structures and versatile physicochemical properties, these materials hold broad application potential in energy, electronics, and catalysis.[2-3] Compared with their three-dimensional counterparts, 2D materials exhibit superior optical performance, electronic transport efficiency, and surface chemical reactivity.[4] These exceptional properties support their role in the development of next-generation high-performance devices.

Recent advances in artificial intelligence have driven the application of generative models in material discovery. Among these, diffusion models,[5] as an emerging class of generative frameworks, leverage their iterative generation approach to simulate complex material morphologies. Owing to their demonstrated effectiveness, diffusion models are among the most promising tools in the field of materials exploration.[6-10] Notable examples include the CDVAE of Tian Xie et al.,[11] which was proposed in 2022 as a diffusion-based framework to generate diverse inorganic materials. Building on this foundation, the subsequent MatterGen model enhanced the ability to design materials with tailored chemical and physical properties.[12] Additionally, Peder Lyngby et al. utilized CDVAE to explore 2D material generation, successfully identifying several 2D structures.[13]

Space group symmetry plays a pivotal role in stabilizing lattice structures and modulating electronic properties, which directly impact material performance in various applications.[14-17] Research has also integrated space group symmetry into generative frameworks to enhance material discovery. Notable works include the DiffCSP of Jiao R et al.,[18] which combines lattice parameters and atomic fractional coordinates in an equivariant diffusion process to predict crystal structures. Similarly, the CrystalFormer of Lei Wang et al. employs an autoregressive transformer to generate crystalline materials with space group control.[19] Other examples include SyMat,[20] which uses variational autoencoders to discover materials with specific properties by learning periodic symmetry. WyCryst,[21] integrates symmetry awareness into the VAE framework to generate inorganic materials.

These studies underscore the transformative potential of symmetry-constrained generative models in materials discovery. However, challenges remain, such as the limited use of symmetry constraints in existing generative models for 2D materials.[22-23] Addressing this challenge deepens our understanding of the role of symmetry in enhancing stability and diversity, advancing high-performance 2D materials.

In this study, we introduce a symmetry-constrained diffusion model (SCDM) that integrates Wyckoff positions within space groups to guide structure generation. Using SCDM, we generated 2000 samples and calculated their convex hull energies via density functional theory (DFT).[24] Following structural relaxation and energy screening, 843 potential stable candidates were identified, none of which are found in existing datasets, including six 2D materials that exhibit phonon spectrum stability.

To benchmark the effectiveness of SCDM, we trained a model without symmetry constraints (UnSCDM) for comparison and evaluated both models via a crystal structure prediction (CSP) model proposed by Lai Wei et al..[25] Further analysis of the generated samples revealed the critical role of symmetry constraints in improving material stability and structural diversity. Additionally, we conducted a detailed investigation of the structures and electronic properties of the candidate 2D materials, confirming their potential for practical applications.

## 2. Approach

Figure 1 provides an overview of the workflow employed to discover 2D materials via the SCDM. First, the crystal structures from open databases of 2D materials undergo symmetry encoding, as depicted in Figure 1a (see Section 2.2 for further details). Next, the symmetry data serve as input for training and generating samples through a diffusion modeling process in Figure 1b (see Section 2.3 for further details). Finally, the generated candidate materials are screened and analyzed to assess the model's effectiveness and novelty, as shown in Figure 1c (see Section 3 for further details).

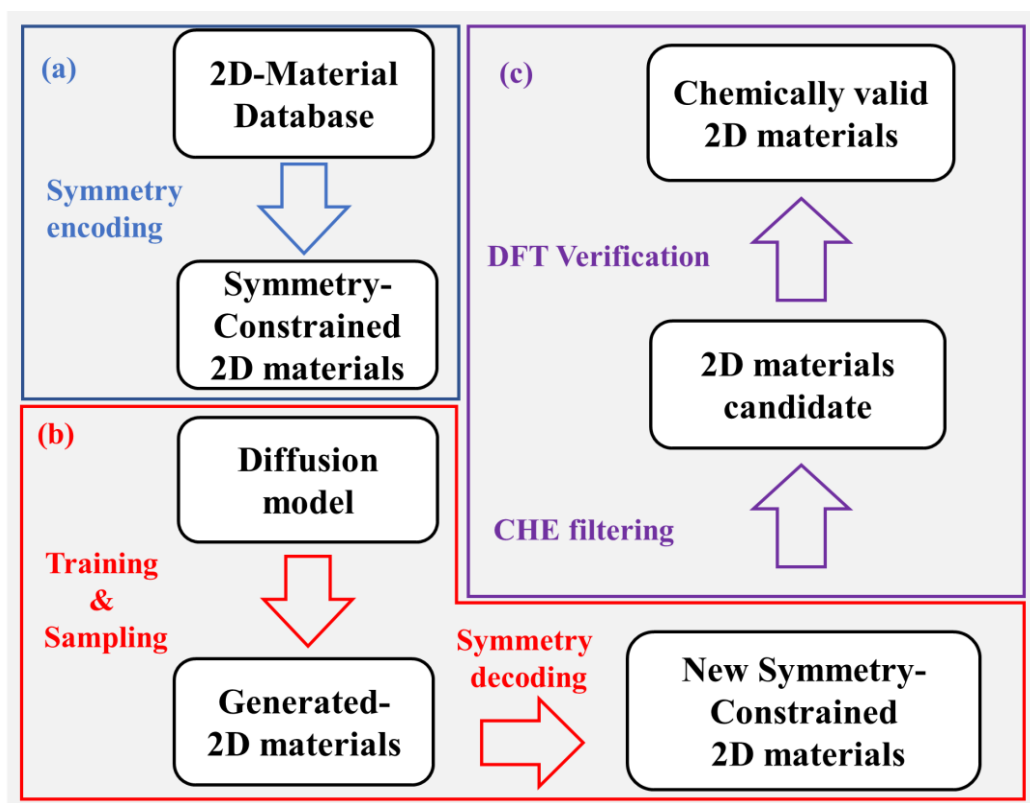

**Figure 1. Schematic of the SCDM workflow.** (a) Symmetry encoding of input crystal structures through designating the corresponding Wyckoff positions; (b) training and sampling of the SCDM, followed by decoding of symmetry information into candidate structures to obtain readable generated 2D materials; (c) screening on the basis of convex hull energy (CHE) and validation via DFT, a filtration process for model assessment.

### 2.1 Data Sets

As summarized in Table 1, the datasets used in this study are drawn from two prominent open 2D material databases: 2dMatPedia[26] and C2DB[27], comprising a total of 10,389 materials. 2dMatPedia, which builds on the Materials Project, includes samples obtained by substituting elements in known 2D materials and by exfoliating bulk 3D materials. C2DB generates samples by decorating established crystal structure prototypes with chemically plausible atomic arrangements. To prepare the datasets for model training, we performed curation to eliminate duplicate entries. Additionally, we excluded structures with unit cells containing more than 40 atoms to ensure compatibility with the model's input constraints. After this refinement, a final set of 5,802 2D materials was selected for training.

**Table 1. 2D material datasets**

| Datasets | Amount | From |
|---|---|---|
| 2dMatPedia[26] | 6351 | Material Project |
| C2DB[27] | 4038 | Known 2D crystal structure prototype |

**2.2 Representation of symmetry-constrained structures**

A critical component of this study is the symmetry encoding of material structures within the dataset, as depicted in Figure 2. For each structure represented by a CIF file, we extract information pertaining to element types, lattice constants, fractional coordinates, and space group numbers, subsequently transforming these features into a matrix representation.

Initially, the element types are represented by their atomic numbers and further encoded via one-hot encoding to distinguish each element. Next, the lattice parameters ($a$, $b$, $c$, $\alpha$, $\beta$ and $\gamma$) are extracted and converted into a lattice matrix via the lattice constant formula (see Equation (1)), enabling a quantitative description of the crystal's geometric configuration. Finally, the fractional coordinates of atoms are directly extracted and encoded into matrices to preserve the spatial arrangement of atomic positions, ensuring the retention of essential structural information.

$$\boldsymbol{L} = \begin{pmatrix} a \\ b \\ c \end{pmatrix} = \begin{pmatrix} a & 0 & 0 \\ b\cos(\gamma) & b\sin(\gamma) & 0 \\ c\cos(\beta) & c\frac{\cos(\alpha)-\cos(\beta)\cos(\gamma)}{\sin(\gamma)} & c\sqrt{1-\cos^2(\beta)-(\frac{\cos(\alpha)-\cos(\beta)\cos(\gamma)}{\sin(\gamma)})^2} \end{pmatrix} \quad (1)$$

The space group numbers are extracted from the CIF files and used to generate the Wyckoff position matrix (Figure 2c), which encodes the spatial arrangement of the fractional coordinates under the constraints of the corresponding space group. Each Wyckoff letter denotes specific symmetry requirements, providing a systematic representation of spatial relationships. Following the symmetry constraint framework established by Cockcroft et al,[28] the Wyckoff positions for space group number 164 (Figure 2c) are categorized as follows: Wyckoff letter '1a' denotes a single unique fractional coordinate, representing the simplest symmetry-constrained position; Wyckoff letter '2d' includes two sets of equivalent fractional coordinates, reducible to a single representative set on the basis of the symmetry operations of the space group; Wyckoff letter '12j' encompasses multiple sets of fractional coordinates, characterizing general positions with minimal symmetry constraints. This systematic encoding process captures fractional coordinates while enforcing space group symmetry constraints, ensuring that the generated crystal structures conform to their specified symmetry properties.

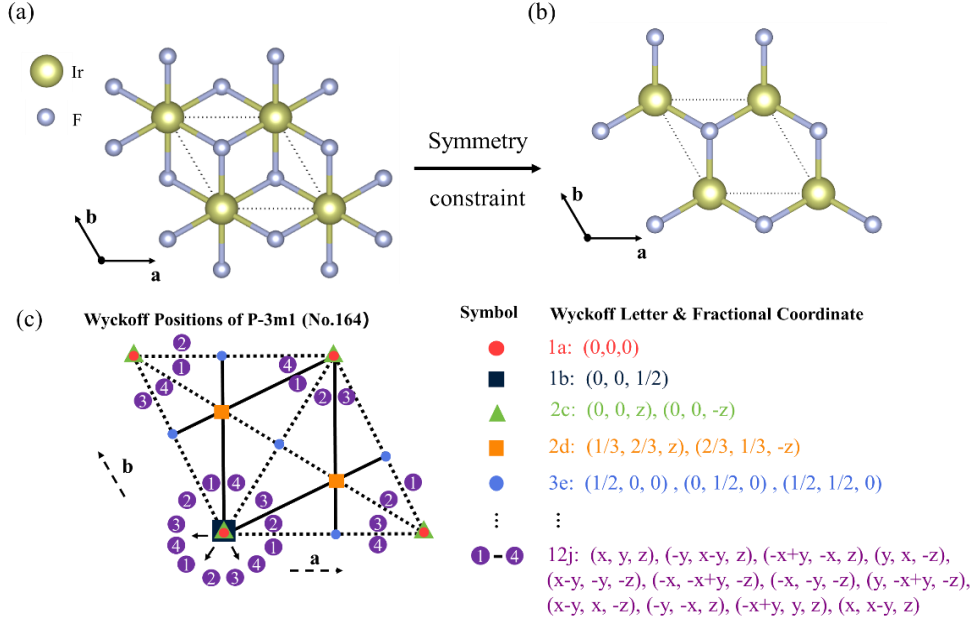

**Figure 2. Symmetry encoding of structures by Wyckoff positions.** (a) The original structure; (b) the structure after applying symmetry constraints; (c) Wyckoff positions of the space group P-3m1. The Wyckoff letters (such as "1a", "1b", etc.) denote fractional coordinates that satisfy the symmetry constraints of the space group. For clarity, a table is provided listing all the fractional coordinates that conform to the symmetry constraints along with their corresponding Wyckoff symbols.

**2.3 SCDM training and symmetry decoding process**

The training process of the SCDM is depicted in Figure 3a. The model leverages the DDPM framework,[5] with inputs consisting of symmetry-encoded data $C_0 = (A_0, L_0, F_0)$ $(t = 0)$. The forward diffusion process is defined as a Markov chain $C_0 \rightarrow \ldots \rightarrow C_T$, where the atom type matrix($A$), lattice matrix ($L$), and fractional coordinate matrix ($F$) independently diffuse via distinct transition kernels, as described in Equation (2):

$$q(A_{t+1}, L_{t+1}, F_{t+1} | A_t, L_t, F_t)$$
$$= q(A_{t+1} | A_t)q(L_{t+1} | L_t)q(F_{t+1} | F_t) \quad (0 \leq t \leq T-1) \quad (2)$$

Here, $(A_T, L_T, F_T)$ denote the atom type matrix, lattice matrix, and fractional coordinate matrix at step t, respectively. The transition kernels $q(A_{t+1} | A_t)$, $q(L_{t+1} | L_t)$, and $q(F_{t+1} | F_t)$ independently diffuse these components. In the reverse sampling process, the model starts from noise-initialized states $(A_T, L_T, F_T)$ and progressively restores the data to its initial state by learning reverse transition kernels, as described in Equation (3):

$$p_\theta(A_{t-1}, L_{t-1}, F_{t-1} | A_t, L_t, F_t)$$
$$= p_\theta(A_{t-1} | A_t)p_\theta(L_{t-1} | L_t)p_\theta(F_{t-1} | F_t) \quad (1 \leq t \leq T) \quad (3)$$

Unlike the forward process, the reverse process involves learning the mean values of the transition kernels, with the variance $\Sigma_\theta(A_t, L_t, F_t)$ being predetermined for each step. The estimation of the mean $\mu_\theta(A_t, L_t, F_t)$ is defined by Equation (4):

$$\mu_\theta(A_t, L_t, F_t) = \frac{1}{\sqrt{\alpha_t}}\left((A_t, L_t, F_t) - \frac{\beta_t}{\sqrt{1-\bar{\alpha}_t}} s_\theta(A_t, L_t, F_t, t)\right) \quad (4)$$

Here, $(A_t, L_t, F_t)$ represents the current diffusion state, whereas $s_\theta(A_t, L_t, F_t, t)$ is the score function predicted by the neural network, which guides the denoising process at step t.

Through the reverse sampling process, the model incrementally reconstructs crystal structures that align with the original data distribution. The Markov chain facilitates stepwise reconstruction of the lattice, atom types, and fractional coordinates, thereby preserving the structural integrity of the original data.

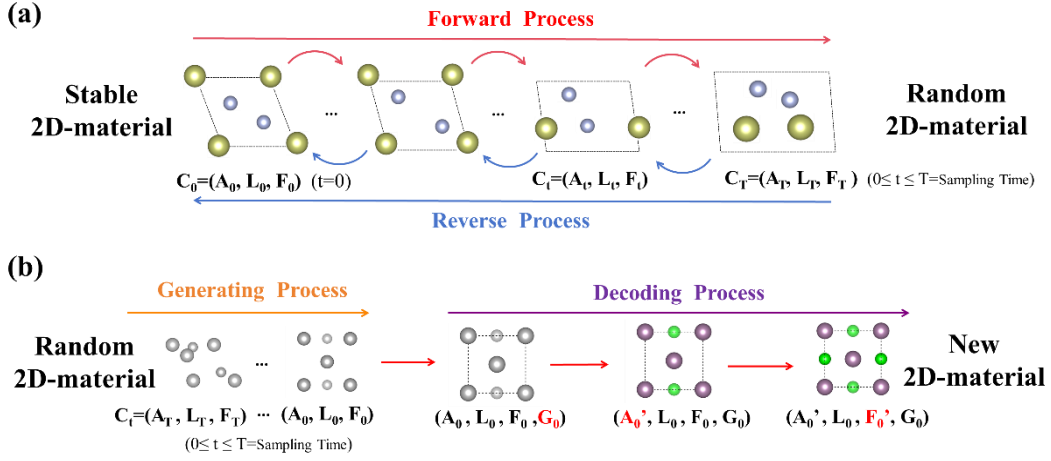

**Figure 3**. **Training, generating, and symmetry decoding process.** (a) The red section shows the forward process during training, where the input structure $C_0 = (A_0, L_0, F_0)(t = 0)$ gradually evolves into a random structure $C_T = (A_T, L_T, F_T)$ over T sampling steps. Here, $A_T, L_T$ and $F_T$ represent the element type matrix, lattice matrix, and fractional coordinate matrix, respectively, at step t. The blue section represents the reverse process, which reconstructs the original structure from the diffused random structure. (b) The orange section shows the generation process, which parallels the reverse diffusion process, reconstructing a stable lattice from a random one. The purple section details the symmetry decoding process, progressively extracting the lattice $L_0$, space group $G_0$, element types $A_0'$, and fractional coordinates $F_0'$ of the generated structure.

Upon completion of the training, we employ SCDM for generation and symmetry decoding, as depicted in Figure 3b. The generated data $C_0 = (A_0, L_0, F_0)$ comprise the atomic type matrix $(A_0)$, lattice matrix $(L_0)$, and fractional coordinate matrix $(F_0)$. These components are decoded to produce crystal structures while preserving symmetry. The decoding process begins with the lattice matrix $(L_0)$, which determines the crystal system and the possible space group $(G_0)$. The one-hot encoding in the atomic type matrix $(A_0)$ is subsequently replaced with the atomic numbers of the elements, producing a new atomic type matrix $(A_0')$. Next, using the space group $(G_0)$ and its corresponding Wyckoff positions, we identify all the fractional coordinates that are consistent with the symmetry constraints imposed by this space group. By incorporating the original fractional coordinates $(F_0)$ into the symmetry-constrained framework, we derive the new fractional coordinates $(F_0')$, which satisfy the symmetry requirements. Ultimately, the decoded structure $C_0' = (A_0', L_0, F_0', G_0)$ satisfies the symmetry requirements of the designated space group, ensuring consistency and physical feasibility.

## 3. Results and Discussion
### 3.1 Evaluating the effectiveness of the SCDM

To systematically evaluate the impact of symmetry constraints on diffusion models, we trained an alternative model, UnSCDM, as a control. Both models followed identical training protocols, with the primary distinction being that UnSCDM does not incorporate space group symmetry constraints during symmetry encoding or structure generation.

For evaluation, we utilized the crystal structure prediction (CSP) model proposed by Lai Wei et al. as the benchmark.[25] Using SCDM and UnSCDM, we reconstructed the original dataset structures and assessed the structural differences before and after reconstruction across multiple CSP metrics. Additionally, we compared the reconstruction results against baseline structures subjected to 15% perturbation in atomic coordinates, as provided by the authors. The evaluation metrics include the energy distance (Figure 4a), sinkhorn distance (Figure 4b), superposition distance (Figure 4c), CrystalNN fingerprint distance (Figure 4d), X-ray diffraction distance (Figure 4e), and orbital field matrix distance (Figure 4f). Box plots of these metrics offer a visual comparison of each model's performance in structure reconstruction.

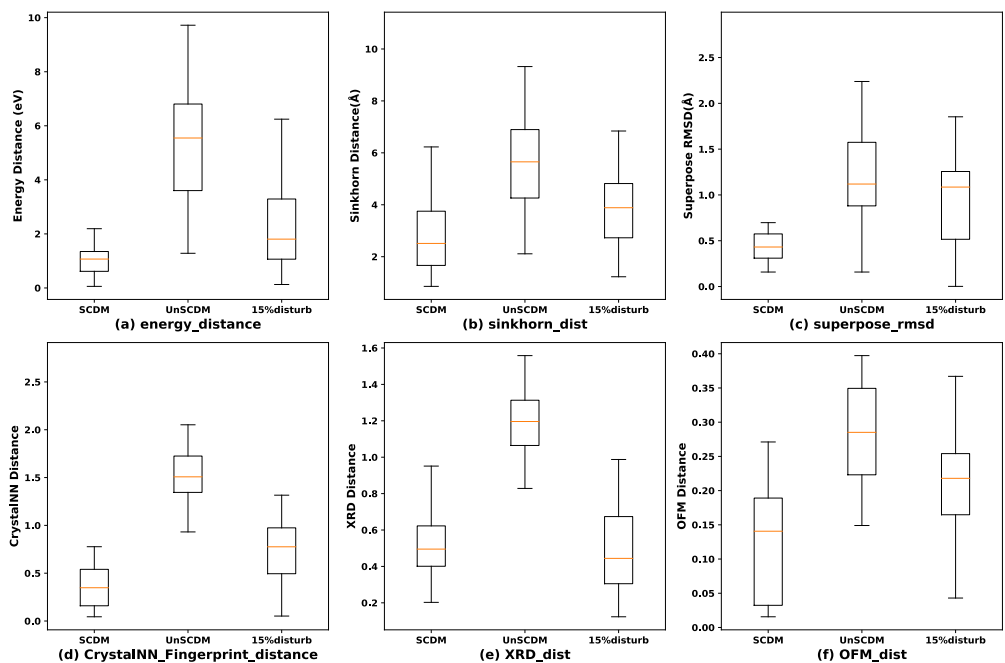

**Figure 4. Evaluation of the reconstruction results via CSP metrics.** The comparison involves structures reconstructed by the SCDM and UnSCDM alongside with original data perturbed with 15% of its atomic coordinates. The metrics include (a) the energy distance; (b) the sinkhorn distance; (c) the superposition distance; (d) the CrystalNN fingerprint distance; (e) the X-ray diffraction distance; and (f) the orbital field matrix distance.

Figure 4a shows that the energy distances of the SCDM-reconstructed structures are lower than that of both the original structures and the control group, indicating closer energy alignment with the original structures. Similarly, Figures 4b and 4c show that the Sinkhorn[29] and Superpose distances for the SCDM-reconstructed structures are smaller. The Sinkhorn distance (SD) compares probability distributions or point clouds in high-dimensional spaces. The superposition distance

(SPD) evaluates the structural similarity between 3D periodic structures. Both metrics indicate better fidelity in fractional coordinate reconstruction for the SCDM group than for the control group. Furthermore, structural similarity metrics, including the CrystalNN fingerprint distance (Figure 4d), X-ray diffraction distance (Figure 4e), and orbital field matrix distance (Figure 4f), consistently show that the SCDM-reconstructed structures deviate less from the original structures. This demonstrates higher structural accuracy and fidelity. Overall, CSP evaluation metrics confirm that diffusion models incorporating symmetry constraints are more effective at reconstruction phase.

**3.2 Evaluation of the Stability and Structural Diversity of the Generated 2D Materials**

To further evaluate the novelty and stability of the generated samples, we performed self-consistent field (SCF) calculations via the Vienna Ab initio Simulation Package (VASP).[30-31] These calculations employed the Perdew-Burke-Ernzerhof (PBE) exchange-correlation functional[32] and the projector augmented-wave method[33] to compute the convex hull energy of candidate materials. According to Aykol et al.,[34] using a threshold of $E_{hull}$ less than 0.2 eV/atom may exclude 26% of known synthesizable polymorphs. Moreover, Peder Lyngby et al.[13] demonstrated that even materials with $E_{hull}$ values exceeding this threshold can be synthesized. Therefore, on the basis of prior studies, we set the screening range to 0.2–0.6 eV/atom. This approach avoids the omission of potentially synthesizable materials while excluding structures with excessively high energy, with a focus on candidates with practical synthesis potential.

The statistical analysis results are shown in Figure 5, which illustrates the distribution of samples meeting the screening criteria and the proportional differences between the samples generated by the SCDM and UnSCDM. In this study, we generated a total of 4000 samples via the two models. First, we conducted a validity check via Pymatgen,[35] which excluded 622 samples with bond lengths exceeding 0.5 Å. Next, the remaining 3,378 samples were analyzed for duplicate structures via the StructureMatcher tool from Pymatgen, ultimately resulting in the retention of 2140 structures. Among these, 1376 (64.29%) were generated by the SCDM model, whereas 764 (35.70%) originated from the UnSCDM model.

We subsequently performed convex hull energy evaluations on these samples. Among the 1145 samples that satisfied the screening criteria, SCDM generated 843 (73.62%) structures, whereas UnSCDM generated 302 (26.37%) structures. These results show that the SCDM symmetry constraints increase the number of candidates within the agreeable range of convex hull energies. Additionally, compared with UnSCDM, SCDM achieved higher validity check pass rates and a greater proportion of samples within the screening range. This demonstrates that symmetry constraints enhance the generation of stable materials and improve the novelty of the generated samples.

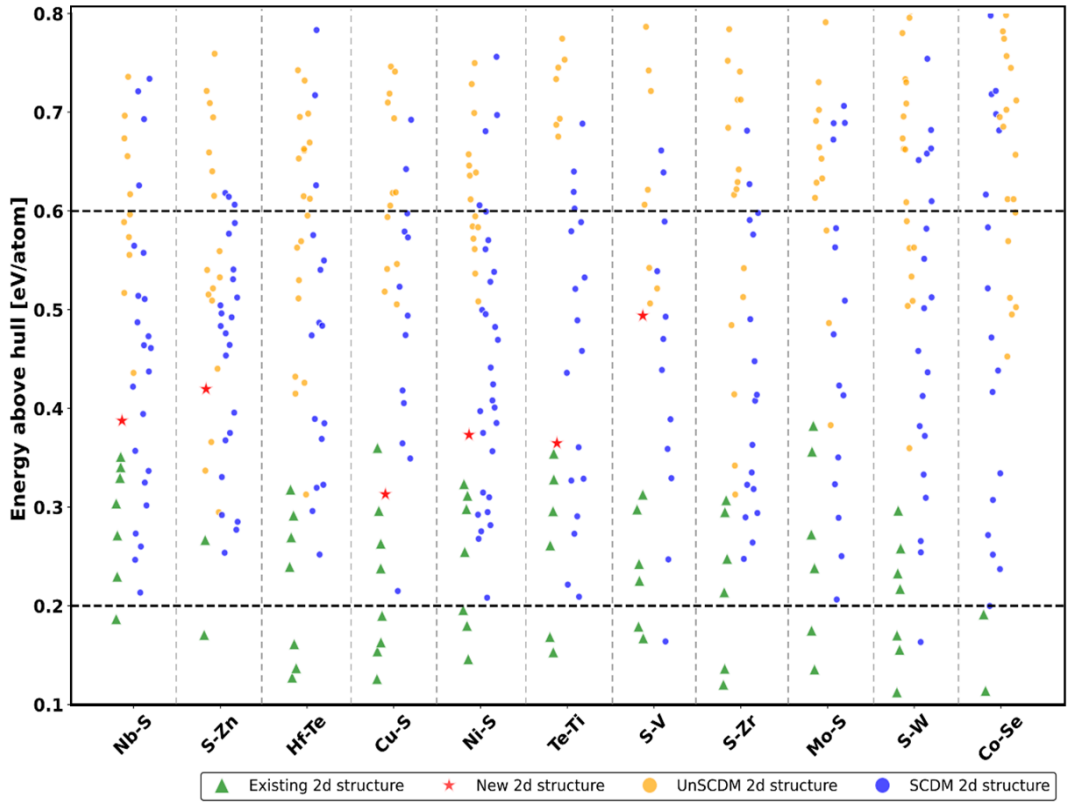

**Figure 5. Distribution of $E_{hull}$ per atom for the generated 2D materials.** Triangles represent stable 2D materials in the database, blue circles represent 2D materials generated by the SCDM, yellow circles represent 2D materials generated by UnSCDM, and pentagrams represent stable 2D materials generated by the SCDM that are not present in the database.

The dynamic stability of candidate materials within the screening range was assessed through phonon spectrum analysis. Six candidates were selected, all originating from SCDM-generated structures. Representative examples of these stable structures are shown in Figure 6. These structures not only meet the convex hull energy criteria but also demonstrate robust dynamic stability in phonon spectrum analysis, highlighting the diversity and potential of SCDM-generated samples for the discovery of 2D materials.

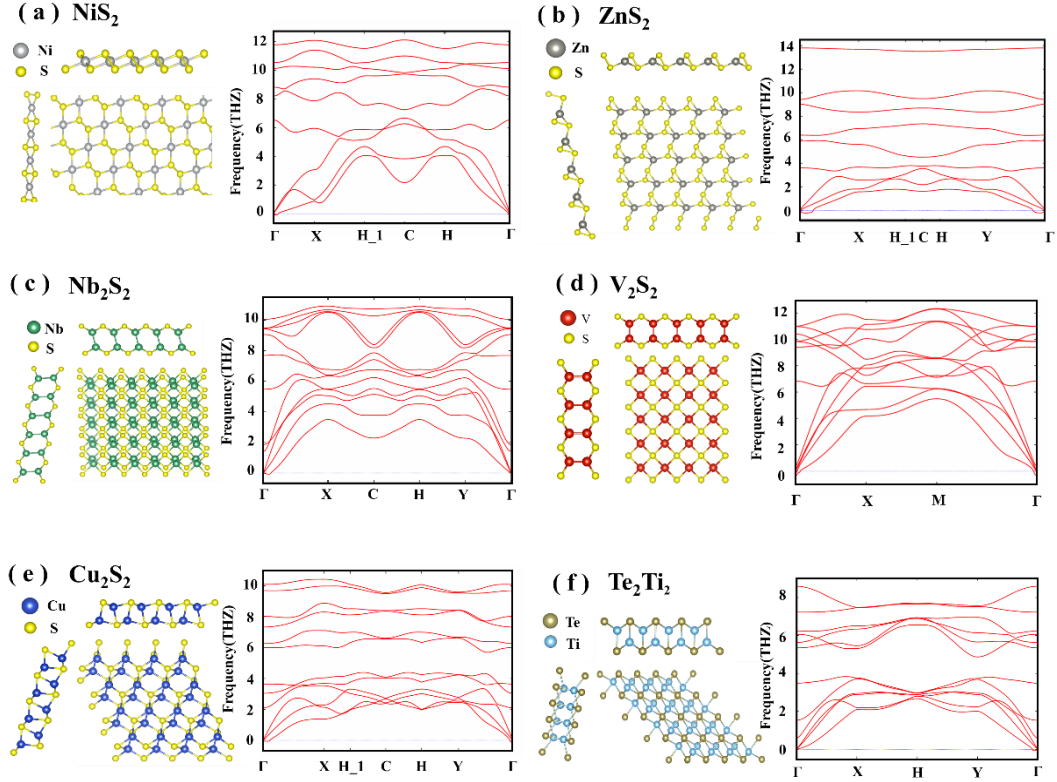

**Figure 6. Structures top and side view with phonon dispersions of 2D crystals (not found in the database).** (a) NiS$_2$, (b) ZnS$_2$, (c) Nb$_2$S$_2$, (d) V$_2$S$_2$, (e) Cu$_2$S$_2$, and (f) Te$_2$Ti$_2$.

**3.3 Structural and Electronic Property Analysis of Candidate 2D Materials**

To explore the potential application prospects of the selected candidate 2D materials, we performed structural relaxation calculations, band structure analyses, and a systematic investigation of their possible uses. Figure 7 shows the band structures and density of states of these materials. Figures 7a-c illustrate the band structure analyses of NiS$_2$,[36] Zn$_2$S,[37] and Nb$_2$S$_2$,[38] which exhibit direct and indirect bandgap characteristics. This behavior highlights their significance in the semiconductor field, particularly for electronic applications, where they demonstrate promising potential. Among these materials, NiS$_2$ serves as a versatile catalyst and finds extensive use in batteries and sensor devices. The wide bandgap of Zn$_2$S enhances its suitability for ultraviolet optical devices. Similarly, the layered structure of Nb$_2$S$_2$ supports its application in optoelectronics and makes it an attractive candidate for energy storage and conversion technologies. Figures 7d-f present the band structures of V$_2$S$_2$, Te$_2$Ti$_2$,[39] and Cu$_2$S$_2$,[36] which display metallic behavior and excellent electrical conductivity. These properties highlight their advantages in terms of efficient electronic conduction and energy transfer. In particular, the metallic nature of V$_2$S$_2$ makes it a strong candidate for high-performance electrical conductivity applications, whereas the exceptional conductivity and distinct crystal structure of Cu$_2$S$_2$ broaden its potential for photocatalysis and other advanced functionalities. Notably, the structural characteristics of these materials have not yet been reported in the literature.

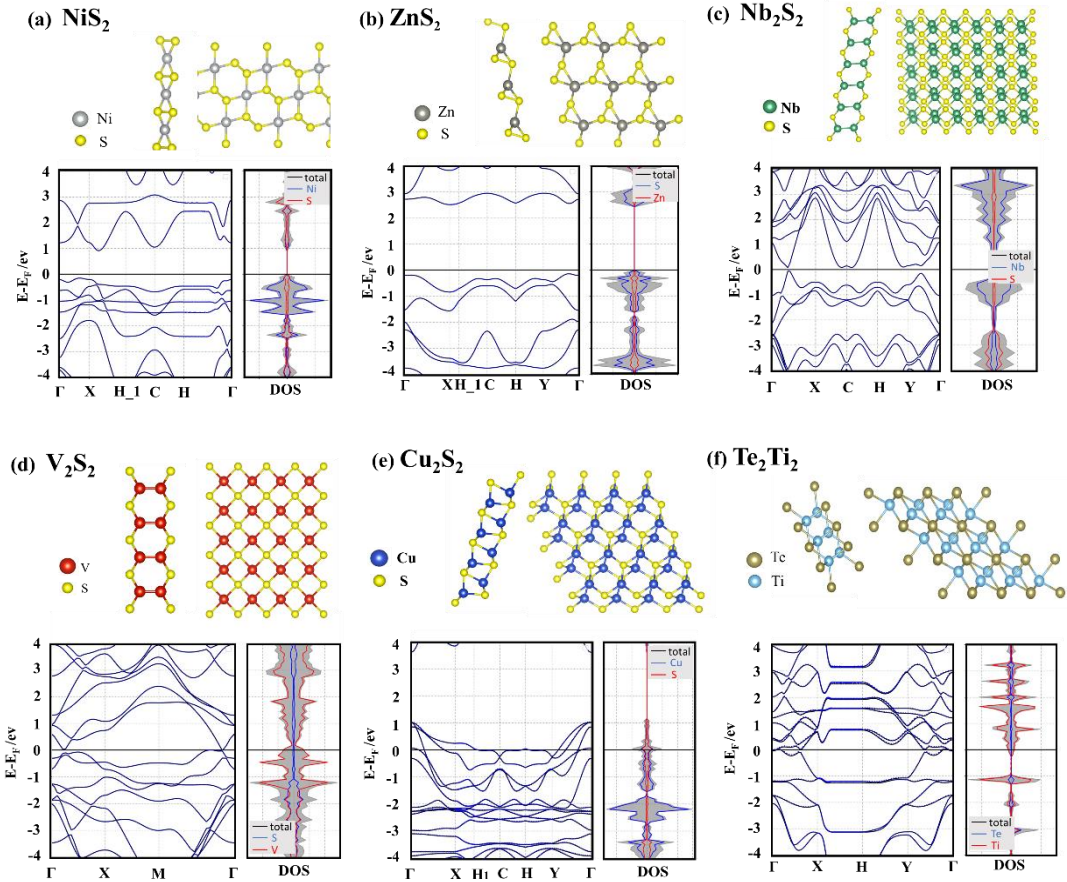

**Figure 7. The corresponding DFT-calculated properties of six 2D materials.** (a) $NiS_2$, (b) $ZnS_2$, (c) $Nb_2S_2$, (d) $V_2S_2$, (e) $Cu_2S_2$, and (f) $Te_2Ti_2$. The band structures and density of states of these materials, illustrate their electronic properties and potential functional applications.

## 4. Conclusion and Future Work

In this study, we developed a SCDM that incorporates spatial group symmetry constraints and demonstrated its effectiveness in generating and screening 2D materials. After training on the 2dMatPedia and C2DB databases, the SCDM successfully generated 2,000 candidate structures, among these, six 2D crystals that have not been reported in the literature were selected for further vibrational properties investigation, all of these selected candidates have exhibit phonon-stable behaviors. A comparative analysis through CSP matrix that join SCDM, UnSCDM, and 15% of perturbed dataset further highlighted the advantages of the SCDM in achieving greater stability and performance. These results emphasize the critical role of symmetry constraints in advancing generative model capabilities. Thus, SCDM offers a transformative approach for discovering of 2D materials. The generated structures were validated through DFT calculations, confirming their reliability and effectiveness. Additionally, its iterative generation process enables the exploration of an extensive structural space, aiding in the identification of 2D materials with substantial potential for several applications. Future efforts could focus on integrating additional constraints related to the physical properties, such as band gaps and magnetism, to improve the stability and functionality of generated structures. These efforts could also be extended to metal-organic frameworks and covalent organic frameworks, making SCDM a promising tool for advancing material design and discovery.

**Acknowledgement**

This work gratefully acknowledges the National Natural Science Foundation of China (No. 11904137, 12074150 and 12174157) and the financial support from Jiangsu University (No. 4111190003).

**Reference**

1. Butler, S. Z.; Hollen, S. M.; Cao, L.; Cui, Y.; Gupta, J. A.; Gutiérrez, H. R.; Heinz, T. F.; Hong, S. S.; Huang, J.; Ismach, A. F., Progress, Challenges, and Opportunities in Two-Dimensional Materials Beyond Graphene. *ACS nano* **2013**, *7*, 2898-2926.
2. Miró, P.; Audiffred, M.; Heine, T., An Atlas of Two-Dimensional Materials. *Chemical Society Reviews* **2014**, *43*, 6537-6554.
3. Chang, C.; Chen, W.; Chen, Y.; Chen, Y.; Chen, Y.; Ding, F.; Fan, C.; Fan, H. J.; Fan, Z.; Gong, C., Recent Progress on Two-Dimensional Materials. *Wuli Huaxue Xuebao/Acta Physico-Chimica Sinica* **2021**.
4. Shanmugam, V.; Mensah, R. A.; Babu, K.; Gawusu, S.; Chanda, A.; Tu, Y.; Neisiany, R. E.; Försth, M.; Sas, G.; Das, O., A Review of the Synthesis, Properties, and Applications of 2d Materials. *Particle & Particle Systems Characterization* **2022**, *39*.
5. Ho, J.; Jain, A.; Abbeel, P., Denoising Diffusion Probabilistic Models. *Advances in neural information processing systems* **2020**, *33*, 6840-6851.
6. Dong, R.; Song, Y.; Siriwardane, E. M.; Hu, J., Discovery of 2d Materials Using Transformer Network‐Based Generative Design. *Advanced Intelligent Systems* **2023**, *5*, 2300141.
7. Chen, C.; Zheng, J.; Chu, C.; Xiao, Q.; He, C.; Fu, X., An Effective Method for Generating Crystal Structures Based on the Variational Autoencoder and the Diffusion Model. *Chinese Chemical Letters* **2024**, 109739.
8. Cao, H.; Tan, C.; Gao, Z.; Xu, Y.; Chen, G.; Heng, P.-A.; Li, S. Z., A Survey on Generative Diffusion Models. *IEEE Transactions on Knowledge and Data Engineering* **2024**.
9. Wines, D.; Xie, T.; Choudhary, K., Inverse Design of Next-Generation Superconductors Using Data-Driven Deep Generative Models. *The Journal of Physical Chemistry Letters* **2023**, *14*, 6630-6638.
10. Li, Z.; Mrad, R.; Jiao, R.; Huang, G.; Shan, J.; Chu, S.; Chen, Y., Generative Design of Crystal Structures by Point Cloud Representations and Diffusion Model. *arXiv preprint arXiv:2401.13192* **2024**.
11. Xie, T.; Fu, X.; Ganea, O.-E.; Barzilay, R.; Jaakkola, T., Crystal Diffusion Variational Autoencoder for Periodic Material Generation. *arXiv preprint arXiv:2110.06197* **2021**.
12. Zeni, C.; Pinsler, R.; Zügner, D.; Fowler, A.; Horton, M.; Fu, X.; Shysheya, S.; Crabbé, J.; Sun, L.; Smith, J., Mattergen: A Generative Model for Inorganic Materials Design. *arXiv preprint arXiv:2312.03687* **2023**.
13. Lyngby, P.; Thygesen, K. S., Data-Driven Discovery of 2d Materials by Deep Generative Models. *npj Computational Materials* **2022**, *8*, 232.
14. Hiller, H., Crystallography and Cohomology of Groups. *The American Mathematical Monthly* **1986**, *93*, 765-779.
15. Cockcroft, J., A Hypertext Book of Crystallographic Space Group Diagrams and Tables. *Acronyms A* **1999**.
16. Urusov, V.; Nadezhina, T., Frequency Distribution and Selection of Space Groups in Inorganic Crystal Chemistry. *Journal of Structural Chemistry* **2009**, *50*, 22-37.
17. Glazer, M.; Burns, G.; Glazer, A. N., *Space Groups for Solid State Scientists*; Elsevier, 2012.

18. Jiao, R.; Huang, W.; Lin, P.; Han, J.; Chen, P.; Lu, Y.; Liu, Y. In *Crystal Structure Prediction by Joint Equivariant Diffusion on Lattices and Fractional Coordinates*, Workshop on"Machine Learning for Materials"ICLR 2023, 2023.
19. Cao, H.; Tan, C.; Gao, Z.; Xu, Y.; Chen, G.; Heng, P.-A.; Li, S. Z. J. I. T. o. K.; Engineering, D., A Survey on Generative Diffusion Models. **2024**.
20. Luo, Y.; Liu, C.; Ji, S., Towards Symmetry-Aware Generation of Periodic Materials. *Advances in Neural Information Processing Systems* **2024**, *36*.
21. Zhu, R.; Nong, W.; Yamazaki, S.; Hippalgaonkar, K., Wycryst: Wyckoff Inorganic Crystal Generator Framework. *Matter* **2024**, *7*, 3469-3488.
22. Liu, Y.; Yang, Z.; Yu, Z.; Liu, Z.; Liu, D.; Lin, H.; Li, M.; Ma, S.; Avdeev, M.; Shi, S., Generative Artificial Intelligence and Its Applications in Materials Science: Current Situation and Future Perspectives. *Journal of Materiomics* **2023**, *9*, 798-816.
23. Yin, H.; Sun, Z.; Wang, Z.; Tang, D.; Pang, C. H.; Yu, X.; Barnard, A. S.; Zhao, H.; Yin, Z., The Data-Intensive Scientific Revolution Occurring Where Two-Dimensional Materials Meet Machine Learning. *Cell Reports Physical Science* **2021**, *2*.
24. Kohn, W.; Sham, L. J., Self-Consistent Equations Including Exchange and Correlation Effects. *Physical review* **1965**, *140*, A1133.
25. Wei, L.; Li, Q.; Omee, S. S.; Hu, J., Towards Quantitative Evaluation of Crystal Structure Prediction Performance. *Computational Materials Science* **2024**, *235*, 112802.
26. Zhou, J.; Shen, L.; Costa, M. D.; Persson, K. A.; Ong, S. P.; Huck, P.; Lu, Y.; Ma, X.; Chen, Y.; Tang, H., 2dmatpedia, an Open Computational Database of Two-Dimensional Materials from Top-Down and Bottom-up Approaches. *Scientific data* **2019**, *6*, 86.
27. Gjerding, M. N.; Taghizadeh, A.; Rasmussen, A.; Ali, S.; Bertoldo, F.; Deilmann, T.; Knøsgaard, N. R.; Kruse, M.; Larsen, A. H.; Manti, S., Recent Progress of the Computational 2d Materials Database (C2db). *2D Materials* **2021**, *8*, 044002.
28. Liu, S.; Li, Y.; Li, Z.; Zheng, Z.; Duan, C.; Ma, Z.-M.; Yaghi, O.; Anandkumar, A.; Borgs, C.; Chayes, J., Symmetry-Informed Geometric Representation for Molecules, Proteins, and Crystalline Materials. *Advances in neural information processing systems* **2024**, *36*.
29. Cuturi, M., Sinkhorn Distances: Lightspeed Computation of Optimal Transport. *Advances in neural information processing systems* **2013**, *26*.
30. Kresse, G.; Hafner, J., Ab Initio Molecular Dynamics for Liquid Metals. *Physical review B* **1993**, *47*, 558.
31. Kresse, G.; Furthmüller, J., Efficient Iterative Schemes for Ab Initio Total-Energy Calculations Using a Plane-Wave Basis Set. *Physical review B* **1996**, *54*, 11169.
32. Perdew, J. P.; Burke, K.; Ernzerhof, M., Generalized Gradient Approximation Made Simple. *Physical review letters* **1996**, *77*, 3865.
33. Blöchl, P. E., Projector Augmented-Wave Method. *Physical review B* **1994**, *50*, 17953.
34. Aykol, M.; Dwaraknath, S. S.; Sun, W.; Persson, K. A., Thermodynamic Limit for Synthesis of Metastable Inorganic Materials. *Science advances* **2018**, *4*, eaaq0148.
35. Ong, S. P.; Richards, W. D.; Jain, A.; Hautier, G.; Kocher, M.; Cholia, S.; Gunter, D.; Chevrier, V. L.; Persson, K. A.; Ceder, G., Python Materials Genomics (Pymatgen): A Robust, Open-Source Python Library for Materials Analysis. *Computational Materials Science* **2013**, *68*, 314-319.
36. An, L.; Li, Y.; Luo, M.; Yin, J.; Zhao, Y. Q.; Xu, C.; Cheng, F.; Yang, Y.; Xi, P.; Guo, S., Atomic-Level Coupled Interfaces and Lattice Distortion on Cus/Nis2 Nanocrystals Boost Oxygen Catalysis for Flexible Zn‑Air Batteries. *Advanced Functional Materials* **2017**, *27*, 1703779.

37. Chakrabarti, A.; Alessandri, E., Syntheses, Properties, and Applications of Zns-Based Nanomaterials. *Applied Nano* **2024**, *5*, 116-142.

38. Ding, X.; Zhang, S.; Zhao, M.; Xiang, Y.; Zhang, K. H.; Zu, X.; Li, S.; Qiao, L., Nbs2: A Promising P-Type Ohmic Contact for Two-Dimensional Materials. *Physical Review Applied* **2019**, *12*, 064061.

39. Liu, H.; Meng, S.; Liu, F., Screening Two-Dimensional Materials with Topological Flat Bands. *Physical Review Materials* **2021**, *5*, 084203.